\documentclass[runningheads]{llncs}
\usepackage{graphicx}
\usepackage{xcolor}
\usepackage{bm}
\RequirePackage{multirow}

\begin{document}

\title{Clustering-based Identification of Precursors of Extreme Events in Chaotic Systems}
\titlerunning{Clustering-based Identification of Extreme Events}
%
\author{Urszula Golyska\inst{1} \and
Nguyen Anh Khoa Doan\inst{1}\orcidID{0000-0002-9890-3173}}
\authorrunning{U. Golyska \& N. A. K. Doan}
%
\institute{Delft University of Technology, Delft 2629HS, Netherlands \\
\email{n.a.k.doan@tudelft.nl}}
\maketitle              
\begin{abstract}
Abrupt and rapid high-amplitude changes in a dynamical system’s states known as extreme events appear in many processes occurring in nature, such as drastic climate patterns, rogue waves, or avalanches. These events often entail catastrophic effects, therefore their description and prediction is of great importance. However, because of their chaotic nature, their modelling represents a great challenge up to this day. The applicability of a data-driven modularity-based clustering technique to identify precursors of rare and extreme events in chaotic systems is here explored. The proposed identification framework based on clustering of system states, probability transition matrices and state space tessellation was developed and tested on two different chaotic systems that exhibit extreme events: the Moehliss-Faisst-Eckhardt model of self-sustained turbulence and the 2D Kolmogorov flow. Both exhibit extreme events in the form of bursts in kinetic energy and dissipation. It is shown that the proposed framework provides a way to identify pathways towards extreme events and predict their occurrence from a probabilistic standpoint. The clustering algorithm correctly identifies the precursor states leading to extreme events and allows for a statistical description of the system’s states and its precursors to extreme events.

\keywords{Machine Learning \and Extreme events \and Clustering \and Chaotic Systems \and Precursors}
\end{abstract}
\section{Introduction}

Extreme events, defined as sudden transient high amplitude change in the system states, are present in many nonlinear dynamic processes that can be observed in everyday life, ranging from rogue waves \cite{rogue_waves}, through drastic climate patterns \cite{climate_effects,climate_change}, to avalanches and even stock market crashes. These kinds of instances are omnipresent and often result in catastrophes with a huge humanitarian and financial impact. As such, the prediction of these events can lead to a timely response, and therefore the prevention of their disastrous consequences.

Unfortunately, many of these problems are so complex that the creation of a descriptive mathematical model is nearly impossible. 
Over the years, three main approaches have been developed in an attempt to tackle this problem.  Statistical methods, such as Large Deviation Theory \cite{large_deviation} or Extreme Value Theory \cite{extreme_value} focus on trying to predict the associated heavy-tail distribution of the probability density function of the observable. Extreme events can be quantified with this method, but are not predicted in a time-accurate manner. Time-accurate forecasting, on the other hand, focuses on predicting the evolution of the system
state before, during, and after extreme events. These methods are often based on Reduced Order Modeling and deep learning to try to accuractely reproduce the dynamics of the systems, including extreme events \cite{farazmand_variational,farazmand_extreme_mechanisms,wan_data-assisted,racca}. Theses approaches generally consider the full system states or a partial physical description which may not always be available. The last approach is precursor identification, which aims at identifying an observable of the system which will indicate the occurrence of extreme events within a specific time horizon. The precursors are typically identified based on characteristics of the system closely related to the quantity of interest which determines extreme incidents. Such precursors have traditionally been identified from physical consideration \cite{kobayashi,murugesan} which made them case-specific and non-generalizable. Recently, a clustering-based method \cite{schmid_analysis,schmid_description} was developed that aims to identify such precursor in a purely data-driven manner by grouping the system's states by their similarity using a parameter called modularity and identifying the pathways to extreme states. This approach showed some potential in providing a purely data-driven approach to this problem of precursor identification which could be generalized to other chaotic systems but was limited to a low-order system.

The purpose of this work is to re-explore this clustering-based approach, apply it to high dimensional systems and assess its predictive capability (statistically and in terms of prediction horizon). The two systems explored here are the Moehlis-Faisst-Eckhardt (MFE) model \cite{mfe} of the self-sustaining process in wall-bounded shear flows, also investigated in \cite{schmid_analysis,schmid_description}, and the Kolmogorov flow, a two dimensional flow which exhibits intermittent bursts of energy and dissipation. To the authors' knowledge, it is the first time this latter system is analysed using such clustering-based approach.

The structure of the paper is as follows. First, the methodology behind the developed algorithm is presented, with the specific steps made to reduce the problem at hand. The criteria for the extreme events and the construction of the transition matrix are described. The concept of modularity is then presented, together with the process of clustering the graph form of the data. Then, in Section \ref{sec:Results}, the results for the two discussed systems are shown, followed by a statistical description of the clusters found by the algorithm. The paper is concluded with a summary of the main findings and directions for future work.

\section{Methodology}
\label{sec:Methods}

\subsection{Preparatory Steps}
The initial step consists in generating a long time evolution of the considered systems. These are obtained using ad-hoc numerical methods which will be discussed in Section \ref{sec:Results} to create appropriate datasets on which to apply the proposed precursor identification technique.

The obtained dataset is then represented in a phase space which is chosen such that the system's dynamics is accurately represented, with the extreme events clearly visible. A combination of the time-dependent states from the data or a modification of them (from physical consideration) is chosen to span the phase space, allowing for a clear representation of the system's trajectory. 

The phase space is then tessellated to reduce the system's size as a given system trajectory (in the phase space) only occupies a limited portion of the full phase space. A singular (high-dimensional) volume resulting from the tessellation of the phase space is referred to as a hypercube. This allows for a precise discretization of the system's trajectory, with no overlaps or gaps.

The time series defining the trajectory in phase space is then analyzed, translating the original data into indices of the respective hypercubes. The result is a time series of the hypercube index in which the system lies at a given time. At this point, the indices of the extreme events are also defined. For a certain point to be considered extreme, all of its quantities of interest must surpass a given specified threshold. These quantities of interest are some (or all) of the parameters spanning the phase space, which are used for defining the extreme events. The criteria for the extreme events are calculated based on user input and are here defined as:
\begin{equation}
\label{extreme_criterion}
    \bigcap\limits_{\alpha=1}^{N_\alpha} |x_\alpha| \geq \mu_\alpha+ d \cdot\sigma_\alpha,
\end{equation}
where $\mu_\alpha$ and $\sigma_\alpha$ are the mean and standard deviation of the considered state parameter $x_\alpha$. The total number of variables used to define the extreme events is $N_\alpha$. Note that $N_\alpha$ could be smaller than the dimension of the phase space as one could use only a subset of the variables of the system to define an extreme event. The constant $d$ is given by the user and is equal for all considered quantities of interest.
All the criteria are then calculated (according to Eq. ~\ref{extreme_criterion}) and applied. A certain region in phase space, for which all of the quantities of interest fulfill the extreme criteria, is labeled as extreme. The tessellated hypercubes which fall into that region are also marked as extreme, which is later used for extreme cluster identification.

\subsection{Transition Probability Matrix and Graph Interpretation}
The tessellated data is then translated into a transition probability matrix. The algorithm calculates elements of the transition probability matrix $\mathbf{P}$ as:
\begin{equation}
   P_{ij} =  \frac{m(B_i \cap \mathcal{F}^{1} (B_j))}{m(B_i)} \;\;\;\;\;\; i,j = 1, ..., N,
\end{equation}
where $P_{ij}$ describes the probability of transitioning from hypercube $B_i$ to hypercube $B_j$, $N$ is the total number of hypercubes on the system trajectory and $\mathcal{F}^{1}$ is the temporal forward operator. The notation $m(B_i)$ represents the number of instances (phase space points) laying in hypercube $i$.


The result of this process is a sparse transition probability matrix $\mathbf{P}$ of the size $M^n$, where $n$ is the number of dimensions of the phase space and $M$ is the number of tessellation sections per dimension.
The $\mathbf{P}$ matrix is generally highly diagonal as for the majority of the time, the trajectory will stay within a given hypercube for multiple consecutive time steps. The transitions to other hypercubes are represented by non-zero off-diagonal elements and are therefore much less frequent.

The transition probability matrix can then be interpreted as a weighted and directed graph. The nodes of the graph are the hypercubes of the tessellated trajectory and the graph edges represent the possible transition between hypercubes. The edge weights are the values of the probabilities of transitioning from one hypercube to another. This representation allows to interpret the system's trajectory as a network and further analyze it as one. This approach also preserves the essential dynamics of the system.

\subsection{Modularity-Based Clustering}
The clustering method adopted for this project is based on modularity maximization, as proposed in \cite{newman_modularity} and applied to identify clusters (also called communities) in the weighted directed graph, represented by the transition probability matrix $\mathbf{P}$ obtained in the previous subsection. 
The Python Modularity Maximization library \cite{modularity_maximization} was re-used here. It implements the methods described in \cite{newman_modularity} and \cite{leicht_community}.

The chosen algorithm is based on maximizing the parameter called \textit{modularity} which is a measure of the strength of the division of a network into communities. This parameter was created with the idea that a good division is defined by fewer than expected edges between communities rather than simply fewer edges. It is the deviation (from a random distribution) of the expected number of edges between communities that makes a division interesting. The more statistically surprising the configuration, the higher modularity it will result in \cite{newman_modularity}.

The modularity value for a division of a directed graph into two communities can be written as:
\begin{equation}
\label{eq:modularity_directed}
    Q  = \frac{1}{m} \sum_{ij} \left ( A_{ij} - \frac{k^{in}_i k^{out}_j}{m} \right ) \delta_{s_i, s_j},
\end{equation}
where $m$ is the total number of edges and $A_{ij}$ is an element of the adjacency matrix and expresses the number of edges between two nodes. In this equation, $s_i$ determines the specific node affiliation, such that when using the Kronecker delta
symbol $\delta_{s_i,s_j}$ only nodes belonging to the same community contribute to the modularity. Notations $k^{in}_i$ and $k^{out}_j$ are the in- and out-degrees of the vertices for an edge going from vertex $j$ to vertex $i$ with probability $\frac{k^{in}_i k^{out}_j}{m}$. A community division that maximizes the modularity $Q$ will be searched for.

The method proposed in \cite{newman_modularity} and used here solves this modularity maximization problem by using an iterative process that continuously divides the network into two communities at each iteration, using the division that increases the modularity the most at that given iteration. The process is finished once no further division increases the overall modularity of the network.

\subsubsection{Matrix deflation}

After the new communities are identified by the algorithm, the transition probability matrix is deflated. A community affiliation matrix $\mathbf{D}$ is created, which associates all the nodes of the original graph to a given community (or cluster). The new transition probability matrix $\mathbf{P_{(1)}}$, which described the dynamics amongst the newly found communities, is then created by deflating the original matrix $\mathbf{P}$ according to:
\begin{equation}
   \mathbf{P_{(1)}} = \mathbf{D^T P D}.
\end{equation}
$\mathbf{P_{(1)}}$ can then again be interpreted as a graph. The process of clustering, deflating the transition probability matrix and interpreting it as a graph is repeated iteratively until one of the two criteria is reached - the maximum number of iterations is exceeded or the number of communities falls below a specified value (making it humanly tractable). After this step, the matrix will still remain strongly diagonal, indicating that most of the transitions remain inside the clusters. Off-diagonal elements will then indicate the transitions to other clusters. 

\subsection{Extreme and Precursor Clusters Identification}
Once the clustering algorithm is done, the extreme clusters are identified. The clusters containing nodes which were flagged as extreme are considered extreme. This is a very general definition, since even clusters containing only one extreme node will be marked as extreme.  The deflated transition probability matrix is then used to identify all the clusters which transition to the extreme clusters. These are considered \textit{precursor clusters}, since they are direct predecessors of the extreme clusters. All the other clusters are considered as \textit{normal clusters} which correspond to a normal state of the system, i.e. far away from the extreme states. Compared to earlier work that also used a similar clustering-based approach \cite{schmid_analysis}, we identify the extreme and precursor clusters and perform a further statistical analysis of them, in terms of time spent in each cluster, its probability of transitioning from a precursor cluster to an extreme one and over which time such a transition will take place. Details on this additional analysis will be provided in Section \ref{sec:Results}.
\section{Results}
\label{sec:Results}
The results of applying the proposed clustering method applied to two chaotic systems exhibiting extreme events, the MFE system, also investigated in \cite{schmid_analysis}, and the Kolmogorov flow are now discussed.

\subsection{MFE system}
The MFE system is a model of the self-sustaining process in wall-bounded shear flows. It is governed by the evolution of nine modal coefficients, $a_i$, whose governing equations can be found in \cite{mfe}. From those nine modal coefficients, the velocity field in the flow can be reconstructed as $\mathbf{u}(\bm{x},t)=\sum_{i=1}^9 a_i(t) \bm{\phi}_i(\bm{x})$ where $\bm{\phi}_i$ are spatial modes whose definitions are provided in \cite{mfe}. The governing equations of the MFE system are solved using a Runge-Kutta 4 scheme and are simulated for a total duration of 100,000 time units (400,000 time steps).

\begin{figure}[!h]
    \centering
    \includegraphics[width = \linewidth]{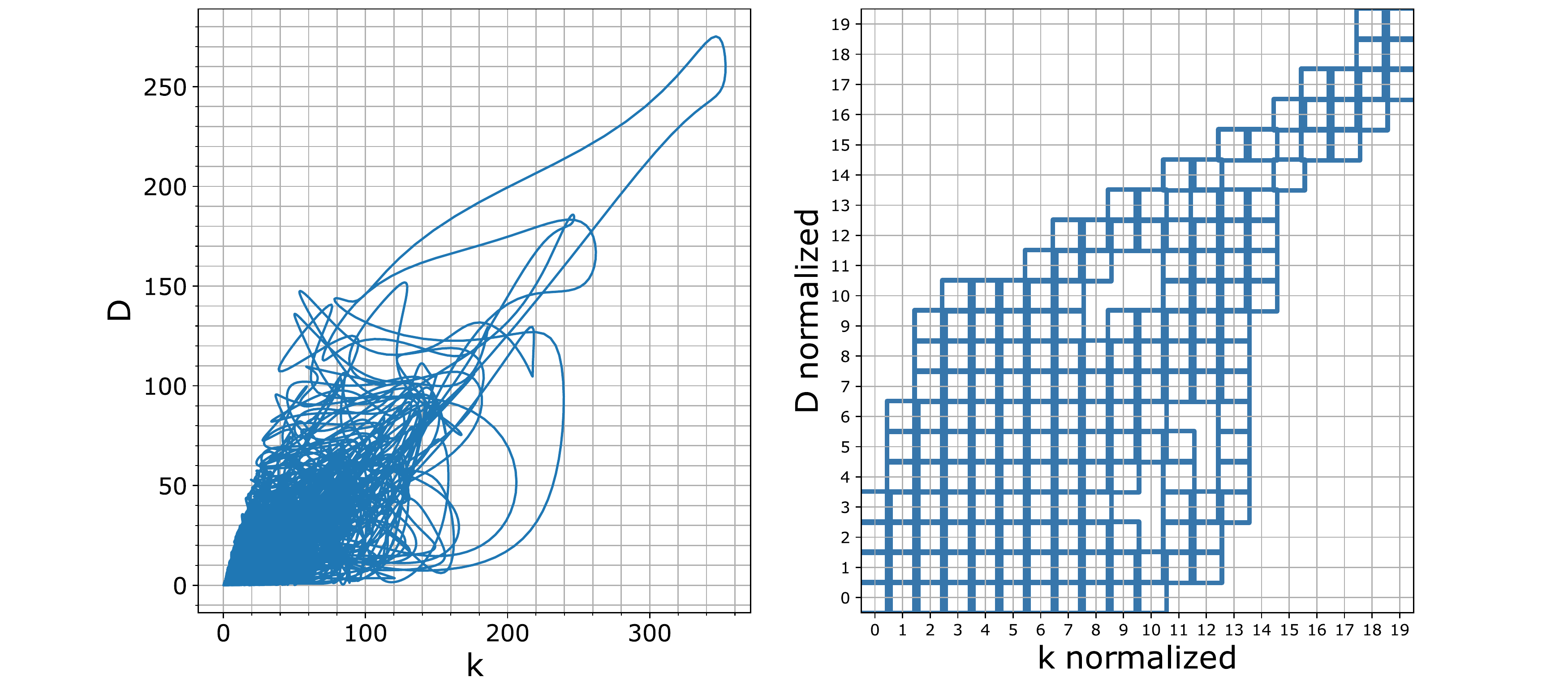}
    \caption{The trajectory of the Moehlis-Faisst-Eckhart model represented in the phase space (left) and in a 20x20 tessellated phase space (right)}
    \label{fig:MFE_phase_space_tess}
\end{figure}

The turbulent kinetic energy $k=0.5 \sum_{i=1}^9 a_i^2$ and energy dissipation $D=tr\left(\nabla \textbf{u} (\nabla \textbf{u} + \nabla \textbf{u}^T)\right)$ were calculated and used as observables. This allowed for a reduction of the size of the system to consider from 9 to 2 dimensions. While such a dimension reduction of the system may seem like a simplification of the problem, it is actually a practical challenge for the precursor identification method as trajectories which were distinct in the full space may now be projected onto a similar reduced space making it harder to appropriately identify a precursor region. The evolution of the system is displayed in the $k$-$D$ phase space in Fig. \ref{fig:MFE_phase_space_tess} (left). The phase space was then tessellated into 20 sections per dimension, which is also seen in Fig. \ref{fig:MFE_phase_space_tess} (right), where $k$ and $D$ are normalized using their respective minimum and maximum and divided into 20 sections. Both $k$ and $D$ are used for clustering as well as the determination of the extreme events. From the evolution shown in Fig. \ref{fig:MFE_phase_space_tess}, the probability transition matrix, $\mathbf{P}$, is constructed as described in Section \ref{sec:Methods} and interpreted as a directed and weighted graph.

\begin{figure}[!h]
    \centering
    \includegraphics[width = 0.44\linewidth]{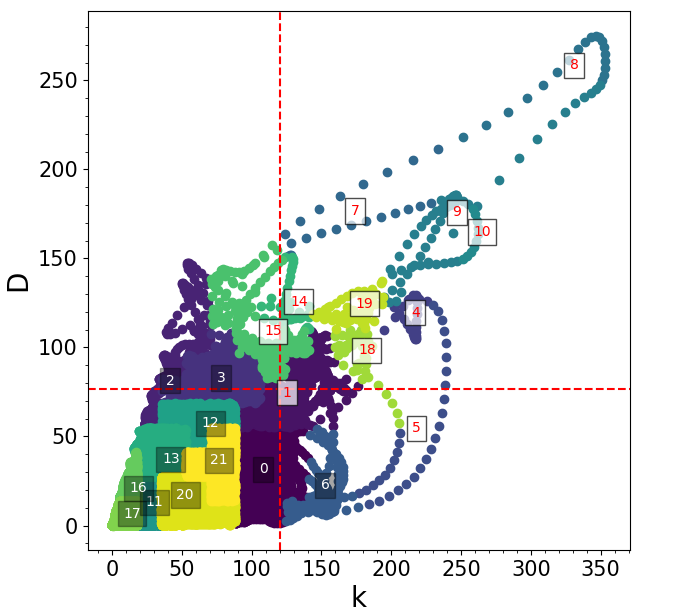}
    \includegraphics[width = 0.44\linewidth]{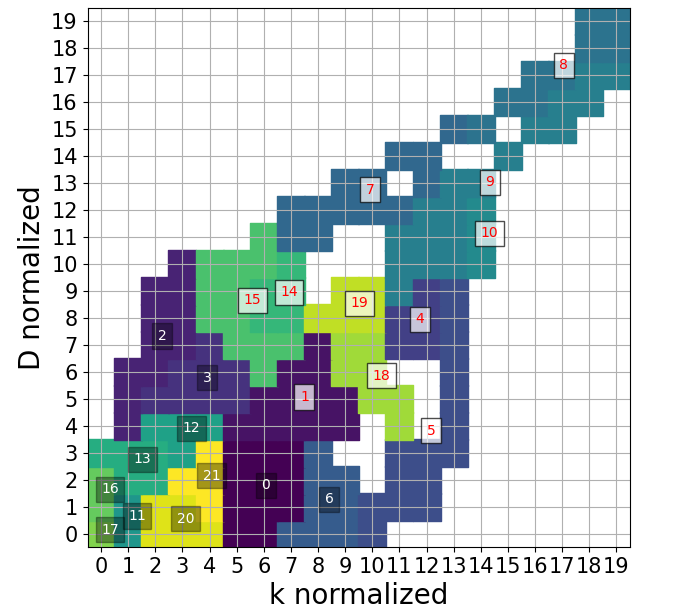}
    \caption{(left) Trajectory of the MFE system in phase space and (right) in a 20x20 tessellated phase space. Red dashed lines represent the region considered as extreme. The cluster numbers used for their identification are shown in the cluster centers; the extreme cluster numbers are marked in red.}
    \label{fig:MFE_clustered}
\end{figure}

The resulting graph consisted of 165 nodes and 585 edges. This was reduced to 27 nodes and 121 edges after the deflation step, which resulted in a deflation of the transition probability matrix from $400 \times 400$ to
$27 \times 27$, a decrease by a factor of over 200. The clustering algorithm identified 27 clusters, out of which 14 were extreme, 6 were precursor clusters and 7 were considered normal state clusters. The clustered trajectory of the MFE system, in tessellated and non-tessellated phase space, is displayed in Fig. \ref{fig:MFE_clustered}, where each cluster is represented by a different color, and the cluster numbers are displayed in their centers. 
In the tessellated phase space (Fig. \ref{fig:MFE_clustered} (right)), the criteria of the extreme event classification are also shown as red dashed lines. These criteria had to be confirmed visually, ensuring that an adequate part of the trajectory lies within them. The majority of the trajectory is located in the bottom left corner, in clusters 20, 25 and 26, which represent the normal state experiencing low $k$ and $D$. The closer to the top right extreme section, the less frequent but more severe the events get.

\begin{figure}[!h]
    \centering
    \includegraphics[width=\linewidth]{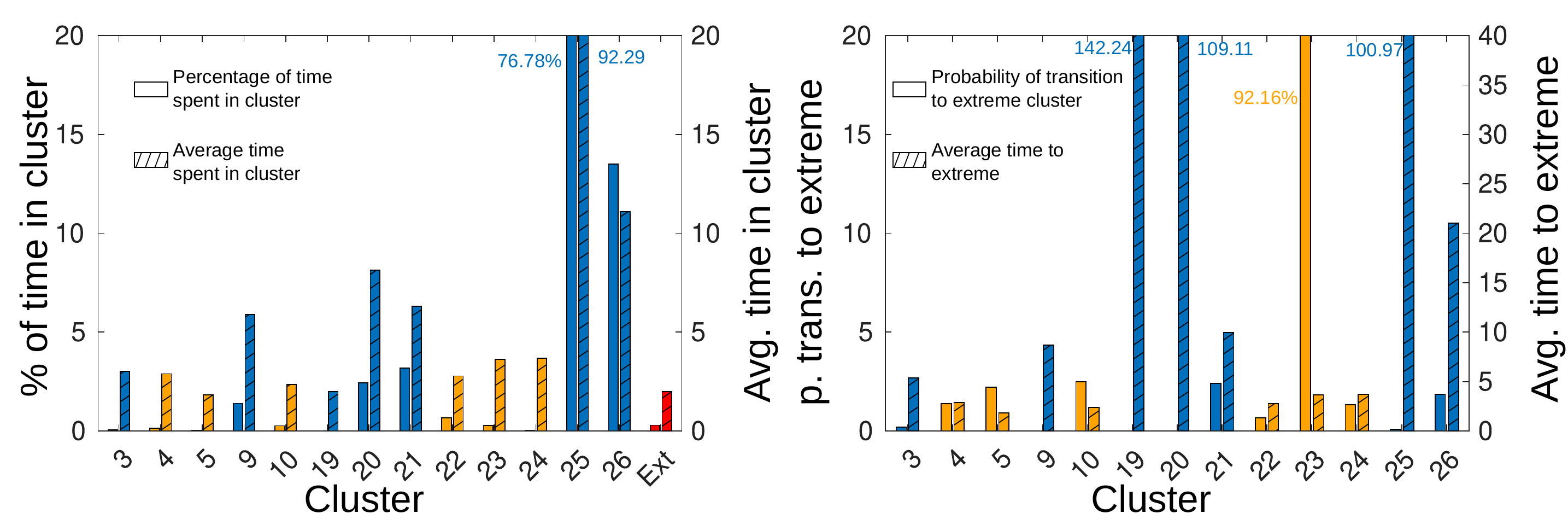}
    \caption{Statistics of the MFE clusters showing (left) the percentage of the time series and average time spent in cluster and (right) the maximum  probability of transitioning to an extreme event and the minimum average time to an extreme event. The extreme clusters are colored red and their direct precursors orange.}
    \label{fig:MFE_stat}
\end{figure}

In Fig. \ref{fig:MFE_stat}, the statistics of the most important clusters are provided. Namely, we compute: (i) the average time spent in each cluster; (ii) the percentage of time spent in each cluster (in proportion to the entire duration of the dataset); (iii) the probability of transition from the given cluster to an extreme one; and (iv) the average time for such a transition to occur from the given cluster. In Fig. \ref{fig:MFE_stat} (left) where the first two quantities are plotted, we can observe that there are a few \textit{normal} clusters (clusters 20, 21, 25 and 26 in blue) where the system spends the largest portion of the time (in average or as a proportion of the total time series duration) while for the precursor (orange) and extreme (red) clusters the proportion is much smaller. This confirms that the proposed approach is efficient in determining clusters which contain the essential dynamics of the system and furthermore, the obtained division helps to contain most of the normal states in one (or few) cluster(s) and isolate the extreme and precursor states.

In Fig. \ref{fig:MFE_stat} (right), among the precursor clusters (in orange), cluster 23 stands out as having a very large probability of transitioning towards an extreme cluster indicating that this is a very critical cluster as, if the system enters it, it will nearly certainly results in an extreme event. For the other precursor clusters, the probability is not as high indicating that, while they are precursor clusters, entering in those does not necessarily results in the occurrence of an extreme event as it is possible that the system will transition again towards a normal cluster. An additional metric shown in Fig. \ref{fig:MFE_stat} (right) is the average time before the transition to an extreme event which is a proxy for the time horizon which would be available to prevent the occurrence of the extreme events. As could be expected, this time horizon is low for precursor clusters (2-3 time units) while it is much larger for the normal cluster (up to 142 time units for cluster 19). The results discussed here highlight how the proposed algorithm provides clusters of system states which can be further classified using a combination of the probability of transitioning towards an extreme cluster and the time horizon over which it can happen. This latter would enable an assessment of the criticality of the system state and cluster.


\begin{figure}[!h]
    \centering
    \includegraphics[width=\linewidth]{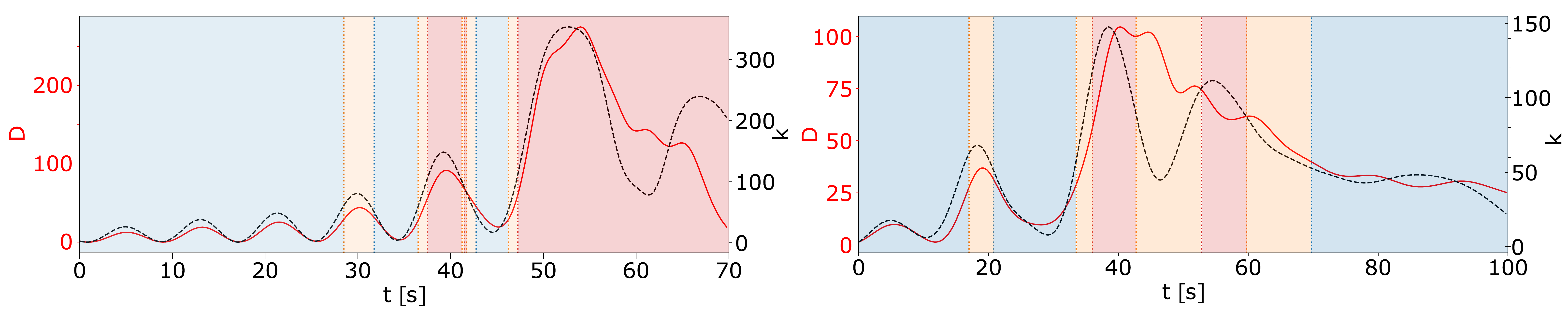}
    \caption{Time series sections of the kinetic energy (dotted black line, right axis) and energy dissipation (red line, left axis) for the MFE system. Background color indicates the type of clusters the system is in (blue: normal; orange: precursor; red: extreme).}
    \label{fig:MFE_peaks_colored}
\end{figure}

To help illustrate the observation made above, the time series of the kinetic energy and dissipation of the MFE system for two extreme events is plotted in Fig. \ref{fig:MFE_peaks_colored} where the plots are colored by the types of cluster the system is in at a given time (normal, precursor or extreme). One can observe that, according to the precursor cluster definition, each extreme cluster is preceded by a precursor cluster. This time during which the system lies in a precursor state varies in the two cases but illustrates the horizon over which actions could be taken to prevent the occurrence of the extreme event.

The false positive rate was also calculated as the ratio of the identified precursors without a following extreme cluster transition to all identified precursor clusters. It showed to be very high, at over 75\%. It was noticed that often not only is there a precursor state preceding, but also following the extreme events, when the trajectory is getting back to the normal states, as can be seen in Fig. \ref{fig:MFE_peaks_colored} (left) around $t=40$. These transitions were accounted for and the corrected rate of false positives was calculated and reduced down to 50\%. Other observations included the fact that extreme events come in batches, often with a false positive slightly preceding it (see Fig. \ref{fig:MFE_peaks_colored} for example where in both cases shown, there is an identified precursor state approximately 10 to 20 time units before the occurrence of the extreme event). This might be leveraged as an additional indication of the upcoming extreme events.

\subsection{Kolmogorov Flow}
The Kolmogorov flow is a two-dimensional flow governed by the incompressible Navier-Stokes equations with the addition of a harmonic volume force in the momentum equation \cite{farazmand_variational}. These are here solved on a domain $\Omega \equiv [0, 2\pi] \times [0,2\pi]$, using a grid of $24 \times 24$ with periodic boundary conditions (same conditions as in \cite{farazmand_variational}), with a forcing term of the shape $f=(\sin(k_fy),0)$, with $k_f=4$ and $Re=40$. The dataset of the Kolmogorov flow is obtained using a pseudo-spectral code and it is generated for a duration of 5,000 time units (500,000 time steps).

The intermittent bursts of energy dissipation, $D$, are of interest for the Kolmogorov flow, with the kinetic energy, $k$, as an observable. In \cite{farazmand_variational}, additional observables related to the extreme events were identified to be the modulus of the three Fourier modes of the velocity field: $a(1,0)$, $a(0,4)$ and $a(1,4)$, which form a triad (where $k_f = 4$ is the forcing wavenumber and where $a(i,j)$ indicates the $i$-th,$j$-th coefficient of the 2D Fourier mode). This approach was applied in the current work, and the algorithm was tested by including one of these three Fourier modes, together with $k$ and $D$, for the clustering analysis, reducing the system dimensionality to 3. This resulted in five different cases, depending on which/how the Fourier mode is considered, which are detailed below:
\begin{itemize}
    \item \textbf{Case 1} - $k$, $D$, and modulus of Fourier mode $|a(1,0)|$
    \item \textbf{Case 2} - $k$, $D$, and absolute value of the real part of Fourier mode $|Re (a(1,0))|$
    \item \textbf{Case 3} - $k$, $D$, and absolute value of the imaginary part of Fourier mode $|Im (a(1,0))|$
    \item \textbf{Case 4} - $k$, $D$, and modulus of Fourier mode $|a(0,4)|$
    \item \textbf{Case 5} - $k$, $D$, and modulus of Fourier mode $|a(1,4)|$
\end{itemize}
A comparative analysis between the different cases was also performed which showed which Fourier mode is the most effective in helping identify precursors of extreme events.
Cases 2 and 3, which include the real and imaginary part of the Fourier mode rather than its modulus, were chosen because of an error encountered for case 1 in the clustering process, which will be detailed hereunder. 
In all cases, the extreme events were defined using only the values of $k$ and $D$.

It should be noted that it would be possible to consider all Fourier modes simultaneously, in addition to $k$ and $D$ which was attempted. However, this resulted in a non-convergence of the modularity maximisation algorithm as it became impossible to solve the underlying eigenvalue problem. It is inferred that this was due to an excessive similarity of the parameters (the various Fourier modes) leading to the underlying eigenvalue problem being too stiff to solve.



\begin{figure}[!h]
    \centering
    \includegraphics[width=\linewidth]{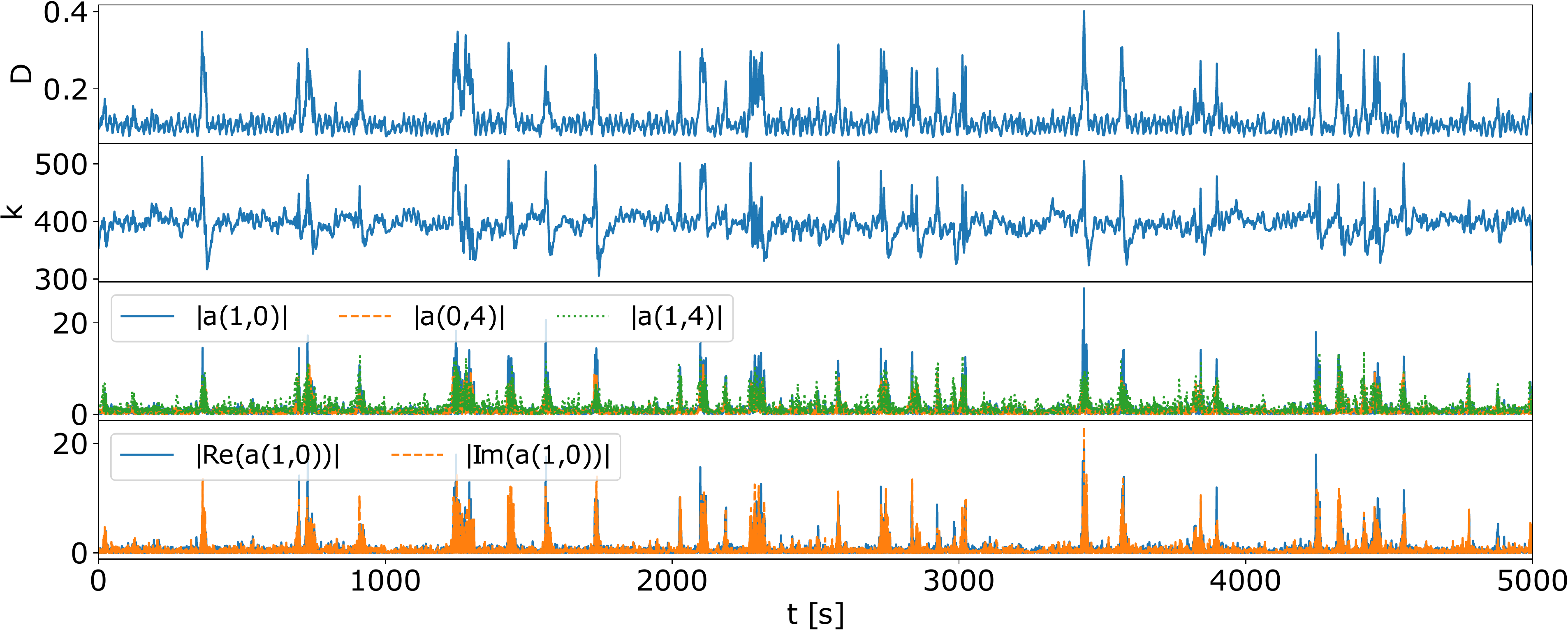}
    \caption{Time series of the energy, energy dissipation and different Fourier modes taken into account for the analysis of the Kolmogorov flow.}
    \label{fig:ts_k_D_fourier_modes}
\end{figure}

The time series of the different parameters taken into account are shown in Fig. \ref{fig:ts_k_D_fourier_modes}. At first glance, all Fourier modes seem aligned, experiencing peak values correlated with the peaks in $k$ and $D$. This could lead to the conclusion that all the cases considered above would lead to similar performance in precursor identification. This was not the case, as will be discussed in more detail below.

\begin{figure}[!h]
    \centering
    \includegraphics[width = 0.94\linewidth]{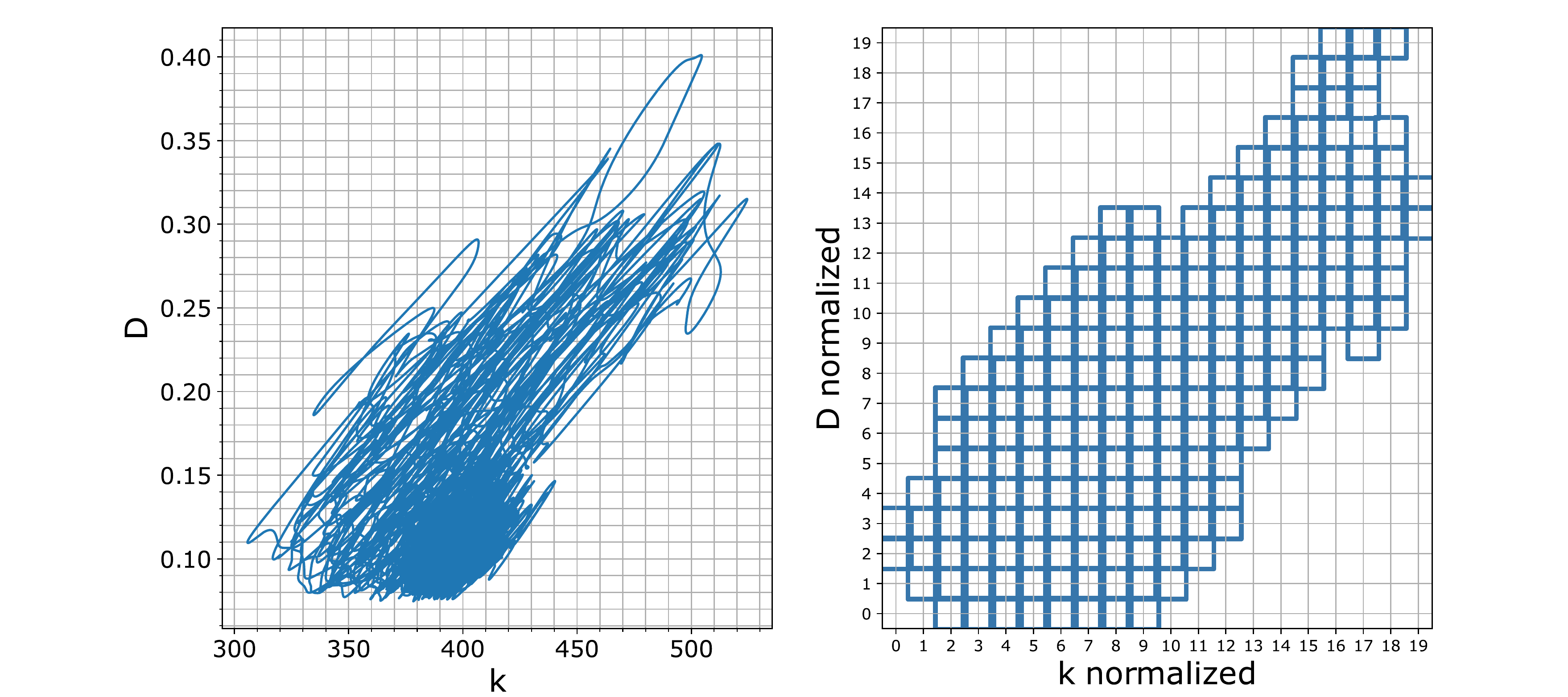}
    \caption{The trajectory of the Kolmogorov flow  represented in the phase space (left) and in a 20x20 tessellated phase space (right).}
    \label{fig:kolmogorov_phase_space_tess}
\end{figure}

The trajectory of the Kolmogorov flow is shown in a phase space spanned by $k$ and $D$, which makes for an easier comparison of the results obtained from different cases, since it is common for all of them. Once again, the tessellation was performed using 20 sections per dimension. The phase space, together with its tessellated form (where $k$ and $D$ are normalized using their respective minimum and maximum and divided into 20 sections), is presented in Fig. \ref{fig:kolmogorov_phase_space_tess}.

\begin{figure}[!h]
    \centering
    \includegraphics[width=0.4\linewidth]{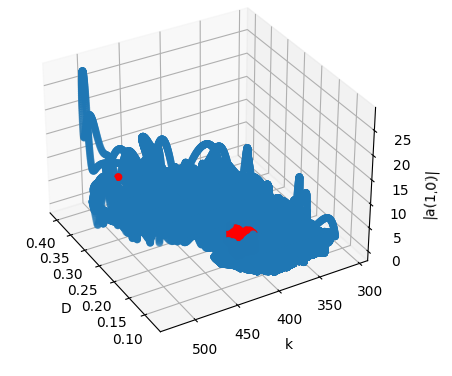}
    \caption{The clustered phase space of the Kolmogorov flow for case 1, with the faulty extreme cluster colored in red.}
    \label{fig:kolmogorov_phase_space_with_error}
\end{figure}

The case including the modulus of $|a(1,0)|$ was tested first (case 1), as this is the one used in \cite{farazmand_variational}. The clustering converged successfully, but resulted in abnormally large values of time spent in both the extreme and precursor clusters. Over a quarter of the total time is spent in precursor clusters. This makes the clustering unusable, since little information can be gained about the probability and time of transitioning to an extreme event. This abnormal result was attributed to a cluster located in the middle of the trajectory, which includes a single node positioned in the extreme section of the phase space (see Fig. \ref{fig:kolmogorov_phase_space_with_error}). 

The other cases yielded much better results, as will be discussed next, although slightly different from each other, varying in the number of clusters and percentage of time spent in extreme and precursor clusters. This highlights the necessity of appropriately choosing the variables used to define the phase space trajectory of the system. 

\subsubsection{Comparison between all cases: }

\begin{figure}[!h]
    \centering
    \includegraphics[width=\linewidth]{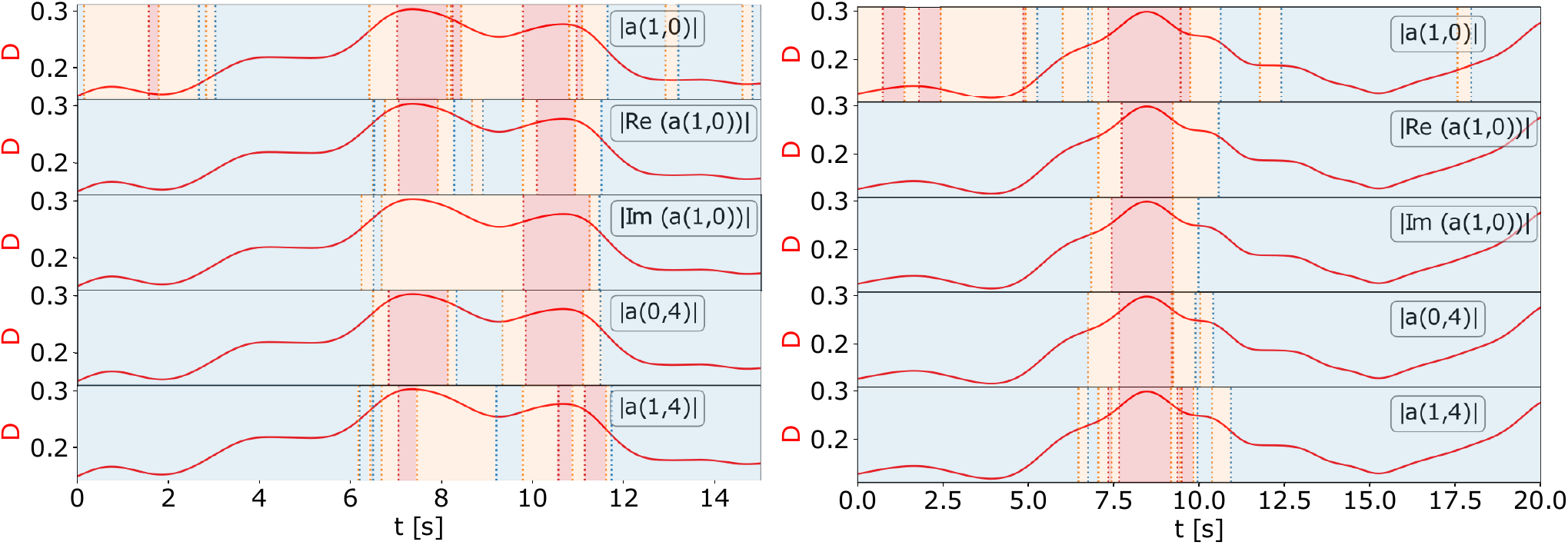}
    \caption{Time series sections for all cases of the energy dissipation, $D$, for the Kolmogorov flow for 2 different extreme events (left and right). Background color indicates the type of clusters the system is in (blue: normal; orange: precursor; red: extreme).}
    \label{fig:kolmogorov_ts_clustered_zoom}
\end{figure}

To understand the differences in the results for the discussed cases of the Kolmogorov flow, segments of the time series of the energy dissipation are plotted with the transitions to different types of clusters marked (Fig. \ref{fig:kolmogorov_ts_clustered_zoom}).
This plot shows a significant difference between the first and other considered cases. At a macroscopic level, cases 2 to 5 seem to identify the peaks of $D$ at the same instances, while for case 1 the transitions are occurring constantly throughout the considered time. This shows the results of incorrectly clustering the system's trajectory making it impossible to deduct any reasonable conclusions and recommendations.

Case 1 shows a very chaotic behavior identifying extreme events where no peaks of $D$ are present. The other cases identify the extreme clusters in a narrower scope, with the smoothest transitions appearing in case 4. 
For the analysis of the extreme events and their precursors, the percentage of time spent in the precursor and extreme clusters, as well as the false positive and false negative rates for all of the discussed cases are presented in Table \ref{tab:kolmogorov_cases_extreme}.

\begin{table}[!h]
\caption{Extreme event statistics of the Kolmogorov flow for different cases.}
    \label{tab:kolmogorov_cases_extreme}
    \centering
    \begin{tabular}{|c|c|c|c|c|c|}\hline
        \textbf{Case} & \multirow{2}{6em}{\textbf{\% time in extreme}} & \multirow{2}{6em}{\textbf{\% time in precursors}} & \multirow{2}{6em}{\textbf{\% false positives}}  & \multirow{3}{6em}{\textbf{\% false positives (corrected)}} & \multirow{2}{6em}{\textbf{\% false negatives}}\\ 
         & & & & &\\
         & & & & &\\ \hline \hline
         1 & 2.78 & 27.85 & 77.15 & 55.11 & 0.0 \\ \hline
         2 & 0.71 & 0.87 & 66.67 & 35.71 & 0.0 \\ \hline
         3 & 0.76 & 1 & 69.44 & 37.96 & 0.0\\ \hline
         4 & 0.85 & 0.79 & 57.30 & 15.73 & 0.0\\ \hline
         5 & 0.81 & 1 & 59.70 & 22.39 & 0.0 \\ \hline
    \end{tabular} 
\end{table}

As mentioned already, case 1 was disregarded, due to faulty clustering resulting in a high percentage of time spent in the precursor clusters. Table \ref{tab:kolmogorov_cases_extreme} also shows that the rate of false positives is the highest for this case, which is explained by the positioning of the faulty extreme cluster (located in the center of the phase space trajectory). All of the discussed cases have a false negative rate of 0 \% which is due to the used definition of precursor cluster as being always followed by an extreme cluster. 
Case number 4, which shows the lowest false positive rates both before and after applying the correction (57.3 \% and 15.73 \% respectively) was chosen for further analysis hereunder.

It should be noted that compared to earlier work \cite{farazmand_variational}, we observe here that using $|a(1,0)|$ did not enable the identification of a precursor of extreme events, while using $|a(0,4)|$ did. This contrasting view may be related to the fact that we use directly the magnitude of the Fourier mode compared to the energy flux between modes used in \cite{farazmand_variational}. Our analysis suggests that when considering the magnitude directly, $|a(0,4)|$ becomes more adapted as is detailed next.

\subsubsection{Case 4:} 
\begin{figure}[!h]
    \centering
    \includegraphics[width = 0.45\linewidth]{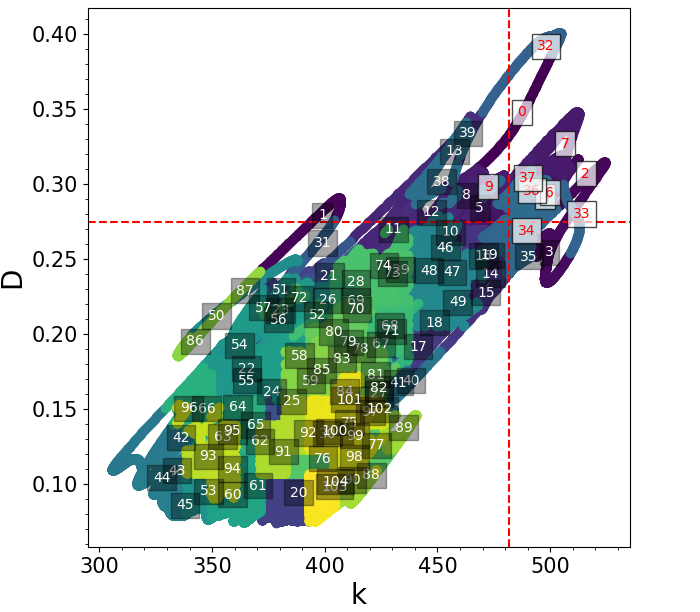}
    \includegraphics[width = 0.45\linewidth]{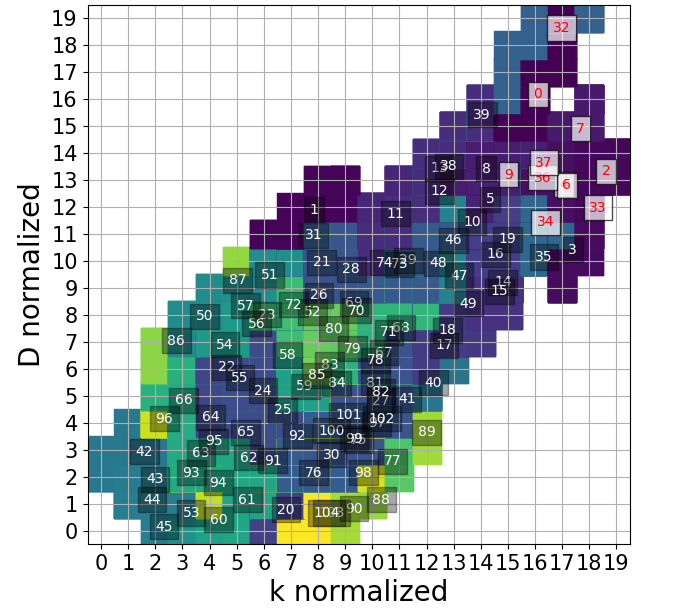}
    \caption{(left) Trajectory of the Kolmogorov flow (case 4) system in phase space and (right) in a 20x20 tessellated phase space. Red dashed lines represent the region considered as extreme. The cluster numbers used for their identification are shown in the cluster centers; the extreme cluster numbers are marked in red.}
    \label{fig:kolmogorov_clustered}
\end{figure}

\begin{figure}[!h]
    \centering
    \includegraphics[width=\linewidth]{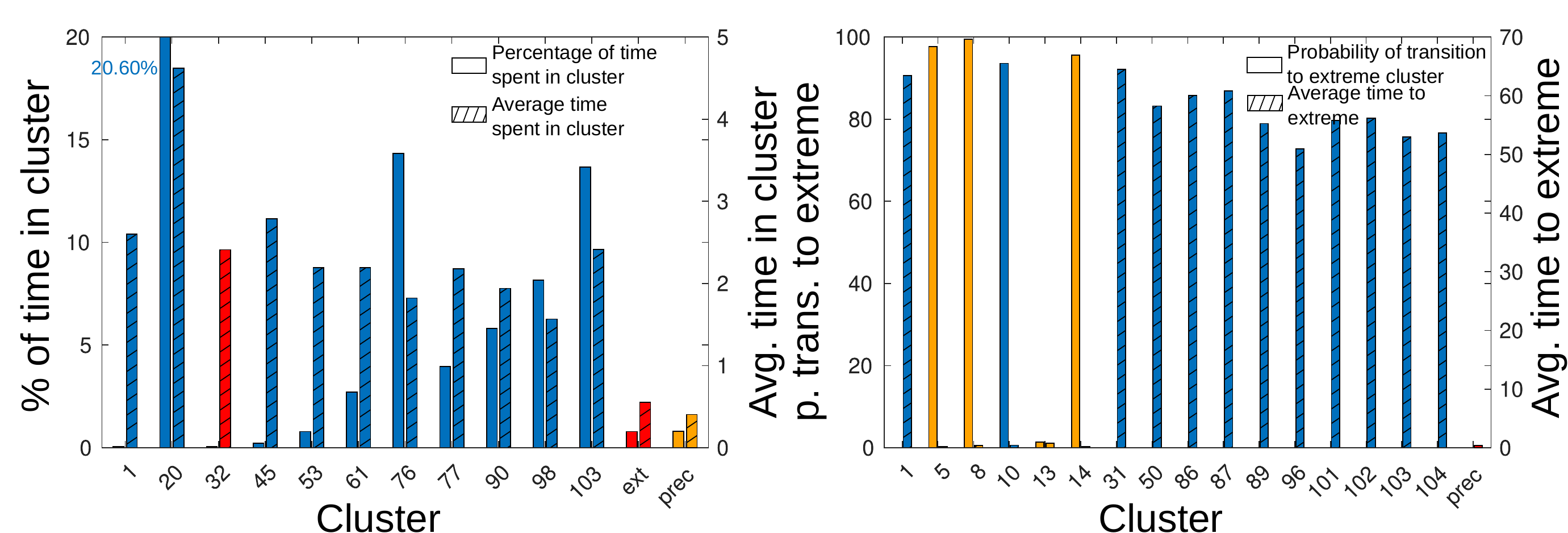}
    \caption{Statistics of the Kolmogorov flow clusters (case 4) showing (left) the percentage of the time series and average time spent in cluster and (right) the maximum probability of transitioning to an extreme event and the minimum average time to an extreme event. Red:  extreme clusters; orange: precursors cluster.}
    \label{fig:kolmogorov_stat}
\end{figure}

The original trajectory of the flow expressed in graph form contained 1,799 nodes and 6,284 edges. After deflation, this decreased to 105 nodes and 719 edges. The transition probability matrix was reduced from 8,000 x 8,000 to 105 x 105. The clustering algorithm identified 105 clusters, out of which 11 were extreme and 8 were precursor clusters. The clustered trajectory of the Kolmogorov flow is shown in Fig. \ref{fig:kolmogorov_clustered}, with each cluster represented by a different color and the cluster numbers displayed in their centers. As for the MFE system, the statistics of the main clusters are shown in Fig. \ref{fig:kolmogorov_stat}.

The resulting statistics showed that most of the time the trajectory is within the normal states, with a few dominating clusters (Fig. \ref{fig:kolmogorov_stat} (left)). There is one extreme cluster (cluster 32) where the system spends on average a large time which is related to a very critical extreme event where the flow exhibit a long dissipation rate (related to a long quasi-laminar state of the Kolmogorov flow). Compared to the MFE system, there is no single (or few) clusters in which the Kolmogorov flow spends the majority of the time which is related to the higher complexity of its dynamics. Regarding the precursor, their combined contribution is shown and, on average, the system spends less than 1 time unit within these precursors states indicating a low predictability horizon. 
A finer analysis is provided in Fig. \ref{fig:kolmogorov_stat} (right) which shows the presence of three precursor clusters with a very high probability of transitioning to an extreme event. Surprisingly, a normal state cluster (cluster 10) also exhibits a high value of probability of transitioning to an extreme event. Because of this, the cluster could also be regarded as another precursor cluster. The average times to extreme event show a large range of values, with low values (below 1 time unit) for the precursor clusters, but reaching up to 64 time units for the clusters considered the safest.

While the cluster identified here may not provide as much prediction time as for the MFE system, the proposed analysis still provided some useful information regarding the system dynamics identifying pathways for extreme events to occur and providing indications of the most critical states that will lead to an extreme events. The statistical analysis that can be carried out thanks to the output of the algorithm reveals the severity of each cluster, which can then be quantified and used for further, more insightful predictions. The clusters can therefore be classified based on any of the calculated parameters to track the real-time progression of the system.

\section{Conclusions}
In this work, we proposed a clustering-based approach to the identification of precursors of extreme events. This was applied to two chaotic systems, the MFE system and, for the first time, the Kolmogorov flow. 
The results showed that the proposed algorithm gives satisfactory results in terms of enabling the identification of such precursors, but has some limitations. The clustering part was successful for most cases, with the essential dynamics of the systems preserved. The results showed that the clusters were found based on the system’s trajectory. 
For the extreme cluster identification, the algorithm managed to correctly identify them, with the exception of one of the cases of the Kolmogorov flow. The clustering-based approach also suggested that a different Fourier mode may be more adapted as a precursor of extreme events compared to the one obtained from a variational-based approach in \cite{farazmand_variational}. The clustering-based precursors were identified correctly and, due to their adopted definition, the false negative rate was equal to zero for all cases. Compared to earlier work \cite{schmid_analysis}, a further statistical analysis of the identified precursor cluster allowed to obtain an understanding of their critical nature, depending on their actual probability of transition to an extreme state and the time horizon over which such transition could occur.

Further work will be dedicated to investigating how to improve the identification and assessment of precursor clusters and applying the proposed method to experimental measurements obtained from systems exhibiting extreme events. Application to real-time data should also be explored, to verify current cluster identification and prediction of its upcoming transitions in such a setting. 

%
%
%
\bibliographystyle{splncs04}
\bibliography{mybibliography}
\end{document}